\documentclass[conference,10pt]{IEEEtran}
\usepackage{float}
\usepackage{comment}
\usepackage[table]{xcolor}
\usepackage{cite}      
\usepackage{graphicx}  
\usepackage{psfrag}    
\usepackage{placeins}
\usepackage{lineno}
\usepackage{url}
\usepackage{subcaption}
\usepackage{caption}
\usepackage{times}
\usepackage{textcomp}
\usepackage{amsmath} 
\usepackage{amsfonts,amsthm,amssymb,mathtools}
\usepackage{balance}
\usepackage[normalem]{ulem}
\usepackage{xcolor}

\usepackage{cuted}
\usepackage{amsmath,bm}
\usepackage{hyperref}
\usepackage{cleveref}
\begin{document}
\title{Synthesis of Through-Wall Micro-Doppler Signatures of Human Motions Using Generative Adversarial Networks}

    
\author{\IEEEauthorblockN{Kainat~Yasmeen and Shobha~Sundar~Ram\\
Indraprastha Institute of Information Technology Delhi, New Delhi 110020 India\\
Email: \{kainaty, shobha\}@iiitd.ac.in}}

\maketitle

\begin{abstract}
Narrowband radar micro-Doppler signatures are heavily used to identify and classify human activities. When the radar is operated in through-wall environments, the complex electromagnetic propagation phenomenology introduces considerable distortions in the micro-Doppler signatures through attenuation and multipath. The problem is particularly severe in inhomogeneous wall scenarios involving multiple wall layers, air gaps, or metal reinforcements. Through-wall radar data collection using simulations and measurements involves significant time and effort. In this paper, we propose an alternative method of synthesizing through-wall radar micro-Doppler signatures from their free space counterparts using the generative adversarial network (GAN). We train the GAN using radar micro-Doppler signatures generated from electromagnetic simulations. We generate the radar data for different human motions, along different orientations, and under diverse through-wall conditions. The synthetic radar micro-Dopplers generated from the neural networks are then evaluated for their realism using a denoising autoencoder, which shows an excellent realism score. 
\end{abstract}
\providecommand{\keywords}[1]{\textbf{\textit{Keywords--}}#1}

\section{Introduction}
\label{sec:Introduction}
Security and surveillance applications use through-wall radar measurements for human detection and activity classification \cite{amin2014through}. A commonly used radar signature for human activity recognition is the micro-Doppler signature that arises from the swinging motions of the human arms and legs \cite{kim2009human}. The micro-Doppler spectrogram is generated by applying the joint time-frequency transforms on the target scattered returns from a narrowband radar \cite{ram2008doppler}. There has been substantial research over the last two decades to classify human activities based on micro-Doppler signatures using different machine learning and deep learning frameworks. \cite{qiao2021human,li2021semisupervised}. Many of these works have tested the proposed algorithms and networks on radar data collected in line-of-sight and homogeneous through-wall scenarios \cite{ram2008through}. Radar signals, propagating through walls, undergo complex electromagnetic phenomenology, including attenuation, ringing, multipath, and refraction \cite{ram2008doppler}. Homogeneous walls with a uniform dielectric constant do not cause considerable distortions to micro-Doppler spectrograms. However, inhomogeneous walls, such as multi-layered walls, walls with air gaps, or metal reinforcements, introduce considerable distortions in the resulting radar signatures \cite{ram2010simulation,vishwakarma2020mitigation}. As a result, micro-Doppler spectrograms distorted by the presence of inhomogeneous walls are often misclassified. Deep neural networks (DNNs) have recently become popular for classifying radar signatures due to the success of DNNs in classifying complex datasets for image processing \cite{gurbuz2020deep}. However, studies of DNNs for micro-Doppler analysis have been severely limited by the difficulty in capturing radar data under diverse through-wall conditions since it requires a lot of time and workforce \cite{jokanovic2016radar}. Training a classification model, particularly a DNN, with restricted training data often results in overfitting \cite{goodfellow2016deep,foster2022generative,nikolenko2021synthetic}.
 
Current research on the classification of radar micro-Doppler data with limited training data can be categorized into three main approaches: In the first approach, the focus is on building robust classifiers that perform well with small training datasets, as demonstrated in \cite{lin2017deep,li2020human}; Second, transfer learning has been employed, wherein knowledge gained from a large-scale dataset is leveraged to assist tasks on a smaller but related dataset \cite{seyfioglu2018dnn}; Lastly, researchers have explored augmenting labeled data gathered from radar measurements or electromagnetic simulations with synthetic data generated from neural networks \cite{mi2018dcgan,erol2020motion}. Among these three approaches, \cite{erol2020motion} reports that synthetic data generation shows superior accuracy and target generalization. In particular, researchers have been exploring the use of generative adversarial networks (GAN) and its variants to synthesize miro-Doppler spectrograms using different frameworks \cite{mi2018dcgan,alnujaim2019generative,alnujaim2021synthesis, erol2020synthesis}. 

GANs are a powerful class of generative modeling tools that use adversarial interactions between two neural networks - a generator and a critic - to produce artificial data identical to actual data \cite{goodfellow2020generative}. In \cite{mi2018dcgan}, deep convolutional GAN (DCGAN) was used to augment the human micro-Doppler spectrograms with a limited set of simulated spectrograms to improve the classification accuracy of the test set. In \cite{erol2020synthesis}, the micro-Doppler signature envelope was integrated as an additional branch in the critic network of a GAN to improve the realism of synthetic data. In \cite{alnujaim2021synthesis}, the authors used the Pix2Pix model, which is conditional GAN (CGAN) for image-to-image translation for synthesizing multi-static spectrograms of human activities from an input spectrogram at a single aspect angle. Our work focuses on synthesizing micro-Doppler spectrograms under complex inhomogeneous through-wall scenarios from input spectrograms generated from free space conditions. Specifically, we propose the GAN-based framework for learning and subsequently incorporating through-wall propagation effects into free space human micro-Doppler spectrograms. 

Ray optical techniques have been commonly used to simulate and analyze through-wall propagation  \cite{le2009ultrawideband}. While these techniques are effective in modeling homogeneous single-layer walls, they are ineffective in modeling propagation phenomenology through inhomogeneous walls with multiple layers, air gaps, or metal reinforcements \cite{dehmollaian2006hybrid}. Instead, full-wave electromagnetic solvers such as finite-difference time-domain (FDTD) simulations have been preferred for modeling the propagation through such complex inhomogeneous structures to achieve higher accuracy. 
In this work, we generate free space human micro-Doppler spectrograms by combining computer animation models of human motions with primitive-based electromagnetic modeling as described in \cite{ram2008simulation}. Subsequently, we generate variations of these spectrograms for different orientations of human motions with respect to the radar. Then, we generate a dataset of corresponding through-wall micro-Doppler spectrograms by hybridizing the human motions in free space with the electromagnetic models of through-wall propagation generated from FDTD as described in \cite{ram2009simulation,ram2010simulation}. The free space and through-wall micro-Doppler spectrograms are provided as training input to the GAN to enable the neural networks to learn how to generate synthetic but realistic through-wall micro-Doppler spectrograms similar to those obtained from electromagnetic simulations. We test the realism of the output through-wall micro-Doppler spectrograms using a denoising autoencoder. Our results show a very good realism score of 1.6.

The remainder of this paper is organized as follows. Section II describes the configuration of the GAN for synthesizing through-wall spectrogram from the free space spectrogram. Section III describes the simulation setup for generating simulated data that are subsequently used for training the GAN. We present the network details and results in Section IV and the conclusion in Section V.
\section{Proposed Methodology}
\label{sec:Methodology}
\begin{figure}[htbp]
   \centering
    \includegraphics[scale=0.35]{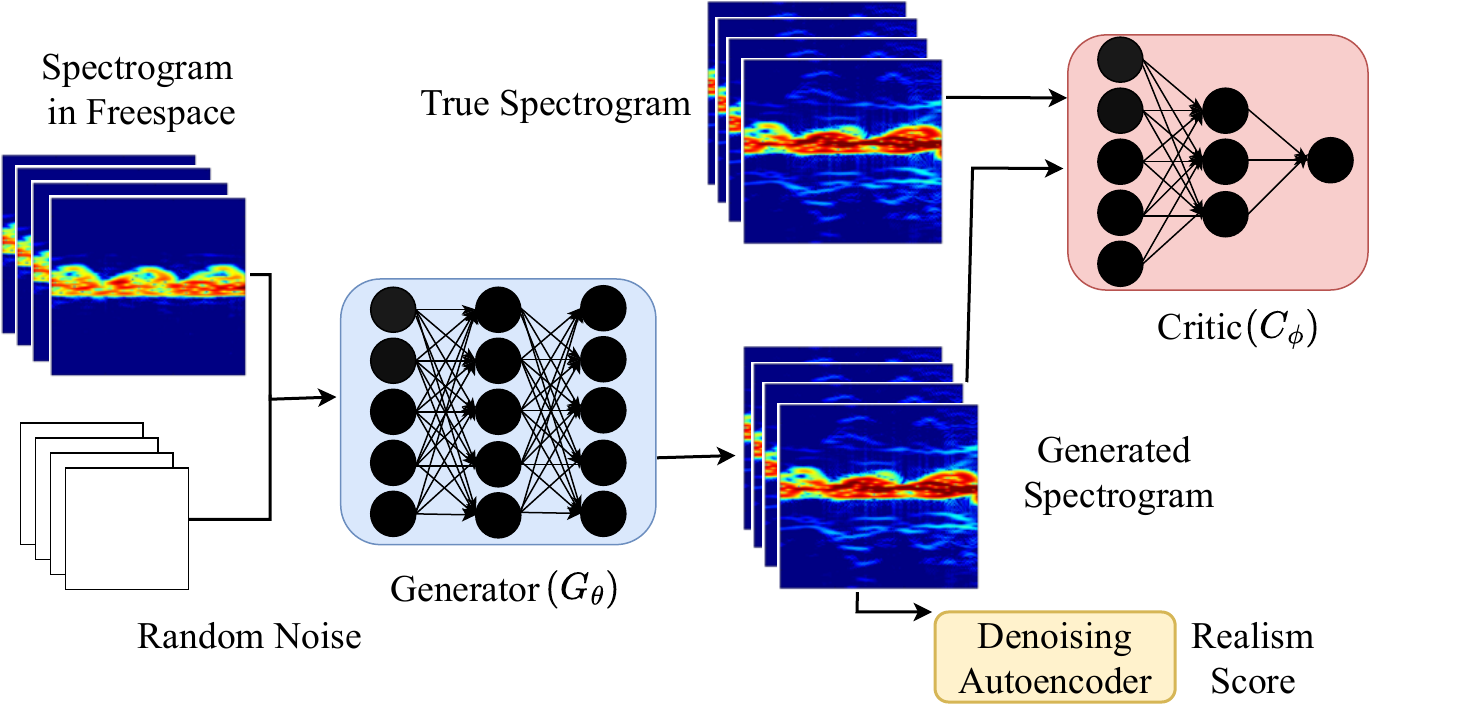}
    \caption{Block diagram of GAN architecture for generating micro-Doppler spectrogram.}
    \label{fig:GAN_model}
\end{figure}

This work aims to synthesize the through-wall micro-Doppler spectrograms from free space micro-Doppler spectrograms using neural networks. Specifically, we use the GAN framework to incorporate the wall effects observed in through-wall micro-Doppler spectrograms into their free-space counterparts. 
GAN comprises two networks known as critic ($C_{\phi}$) with weights $\phi$ and generator ($G_{\theta}$) with weights $\theta$ as shown in the Fig. \ref{fig:GAN_model}. During the training stage, the input to the generator is free space micro-Doppler spectrograms $x$ concatenated with a latent noise vector $\mathcal{N}$ to introduce randomness.
The free space spectrograms are generated from electromagnetic simulations of dynamic human motions, which are described in the following section.
The output of the generator is the corresponding synthetic/fake through-wall micro-Doppler spectrograms $\hat{y}=G_{\theta}(x||\mathcal{N})$.
The free space spectrograms are hybridized with electromagnetic simulations of through-wall propagation methodology to generate through-wall micro-Doppler spectrograms $y$, which are henceforth referred to as the real spectrograms. 
Both the real spectrograms and fake spectrograms are provided as input to the critic network. The generator aims to map the distribution of the generated spectrogram, $\hat{y}$, to the distribution of true spectrograms, $y$. On the other hand, the critic network's objective is to distinguish the real spectrograms from the fake spectrograms. Thus, both networks work in an adversarial manner to facilitate the synthesis of more realistic samples from the generator. This is accomplished using a loss function, $V(G_{\theta}, C_{\phi})$, which is defined as follows:
\begin{multline}
\label{eqn:GAN}
\min_{G_{\theta}} \max_{C_{\phi}} V(G_{\theta},C_{\phi})= \\
\min_{G_{\theta}} \max_{C_{\phi}} \log(C_{\phi}(y)) +\log(1-C_{\phi}(G_{\theta}(x||\mathcal{N})))
\end{multline}

During the test stage, we provide test samples of free space spectrogram $\widehat {x}$ along with random noise to the generator to synthesize the through-wall spectrograms $\widehat{y}$.
To evaluate the performance of the generative architecture, we have considered a realism error that is a popular metric in literature \cite{dhurandhar2018explanations}. Here, we provide the synthetic data as input to a denoising autoencoder (DAE) that is previously trained on the real data collection to provide output (DAE($\widehat(x)$)). The DAE is trained to learn how to distribute real data. During the test, when the DAE is provided with a realistic fake sample, then there is a low reconstruction error, $(||\widehat(x)-DAE(\widehat(x)||_{2}^{2})$, between the DAE's output and the input. On the other hand, if the input sample is unrealistic, then the DAE has a high reconstruction error between the output and input sample. Ideally, during testing, the realism error should be near zero.
\section{Simulation Setup}
\label{sec:Sim}
In this section, we explain the simulation method to generate a database of micro-Doppler spectrograms under free space and through-wall scenarios based on the technique described in \cite{ram2010simulation}.
The free space radar data from dynamic humans are simulated by combining computer animation data and primitive-based electromagnetic models of humans. Then, the through-wall radar data are generated by hybridizing these models of humans with the through-wall propagation models obtained using FDTD techniques. The radar data in free space and through-wall scenarios are then processed using a short-time Fourier transform (STFT) to generate the corresponding micro-Doppler spectrograms.

\subsection{Model of Through-Wall Propagation}
\begin{figure}[htbp]
\centering
\includegraphics[scale=0.5]{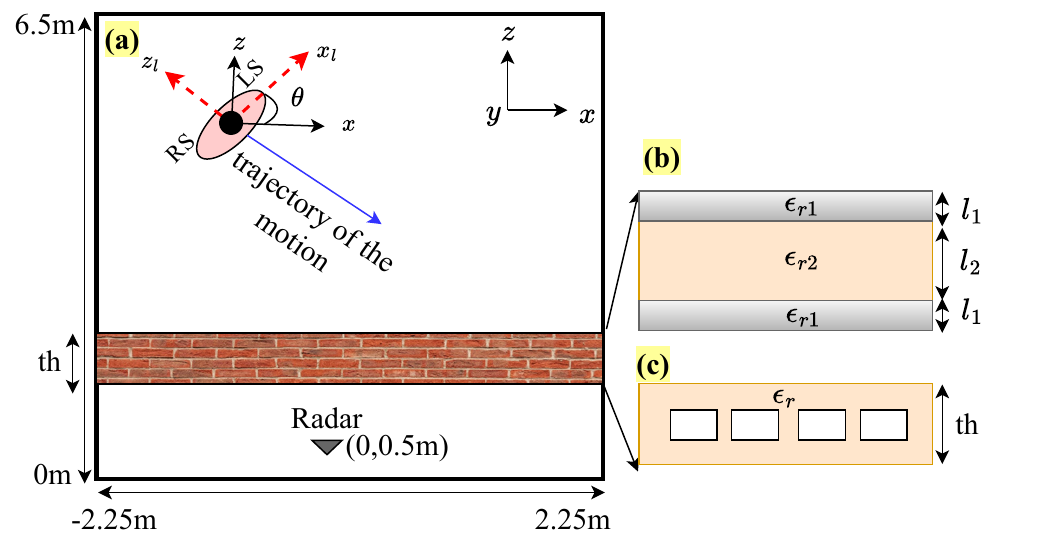}
\caption{a) FDTD simulation setup, (b) a layered wall, and (c) a wall with air gaps.}
\label{fig:simulation_setup}
\end{figure}

We consider a two-dimensional (2D) simulation framework to model through-wall propagation to reduce the overall computational complexity of the problem and because most walls show homogeneity along the height.
We consider a 2D simulation space spanning -2.25 m to 2.25 m along the $x$ axis and 0 m to 6.5 m along the $z$ axis as shown in Fig. \ref{fig:simulation_setup}. Here, the ground is denoted by the $xz$ plane with the height along the $y$ axis. The source excitation is a narrowband sinusoidal line source at 2.4 GHz at $(0,0.5)$ m. The simulation space is enclosed by a perfectly matched layer with a thickness twice the wavelength ($\lambda_c$). The entire 2D space is divided into uniform grids of $\lambda_c/10$ resolution, and the time step is chosen to satisfy the Courant stability conditions. 

We consider two types of inhomogeneous wall scenarios based on conditions encountered in the real world, as shown in Fig.\ref{fig:simulation_setup}. The first wall type consists of three dielectric layers along the $z$ axis, uniform along the $x$ axis. This type of multi-layered wall models real-world dielectric walls with insulation materials/facades on either side. The inner layer has a higher dielectric constant, $\epsilon_{r_2}$, and thickness, $l_2$, than the two outer layers of dielectric constant, $\epsilon_{r_1}$ and thickness, $l_1$. The second type of wall is a homogeneous dielectric wall with relative permittivity, $\epsilon_r$, and thickness, 
 $th$, with periodic air gaps. The length and thickness of the air gaps along $x$ and $z$ are $0.25\times0.1$ m, respectively. Typically, materials used for walls are wood, cement, brick, stone, mud, or cinder. The relative dielectric constants of these materials have been reported to vary from 2 to 8 in literature \cite{balanis2012advanced, allen2019fundamentals}. 
For each through-wall propagation simulation, we consider distinct dielectric constants and thicknesses from the limits specified in Table.\ref{table:total_cases}. In each FDTD simulation, the source excitation generates transverse magnetic (TM) waves propagating in the 2D space based on FDTD equations \cite{yee1966numerical}. 
\begin{table}[!htbp]
\centering
\caption{Dielectric constant and thicknesses of walls}
\label{table:total_cases}
\begin{tabular}{*5c}
\hline 
\noalign{\vskip 1pt}
    {Wall type}&{Parameter}&{Values}&{cases}&{total cases}\\
\hline 
\noalign{\vskip 1pt}
  {Multi-layered wall}&{$\epsilon_{r_2}$}&  {4-8}&{5}&{}\\
  {}&            {$\epsilon_{r_1}$}&  {2-3}& {3}& {}\\
  {}&            {$l_2$}&{15-20 cm}& {2}& {} \\
  {}&            {$l_1$}&{5-10 cm}& {2}&{60}\\\hline
  {Wall with airgaps}&  {$\epsilon_{r}$}& {4-8}&  {10}&{}\\
  {}& {$th$}&  {20-30cm}& {2}&{}\\
  {}& {airgaps}&  {3-5}& {3}&{60}\\  \hline
\end{tabular}
\end{table}
The simulator runs for $65.4$ ns with a time resolution of 0.02 ns. The time-domain electric field at every point $\vec{\rho}$ in the FDTD grid space is then fast Fourier transformed to derive the wall-transmission response $H(f_c,\vec{\rho})$ at carrier frequency $f_c$.
\subsection{Simulation of micro-Dopplers of human motions}
Next, we consider a simulation model of a human moving before the radar, as shown in Fig. \ref{fig:simulation_setup}, over 1.53 seconds. The human is a three-dimensional figure with the height along the $y$ axis. 
The computer animation data, obtained from a motion capture database, has two pieces of information: i) The initial pose of the human skeleton of $B$ bones connected through joints. The root joint is located at the base of the spine, and the other joints are connected in sequence to parent joints through the fixed bone lengths up to the root joint. ii) The subsequent description of the motion of the human through frames. In each frame of the human motion, the root joint is subject to 6 degrees of freedom (DOF) - translation of the position vector of the root along the $x$, $y$, and $z$ axes and Euler rotation angles about the three axes. The remaining joints have only 3 DOF in each frame, which is the Euler rotational angle of the joint. Through translation operation on the root joint, we fix the initial position of the human at a desired starting position within the FDTD room space. Next, through yaw rotation of the joints of the human body, we orient the trajectory motion of the human along the desired aspect with respect to the radar. The desired yaw rotation angle $\theta$ is computed based on the rotation of the local $x_l$ axis formed between the two shoulder joints and $x$ axis as shown in Fig.\ref{fig:simulation_setup}.
\begin{figure}[htbp]
\centering
\includegraphics[scale=0.2]{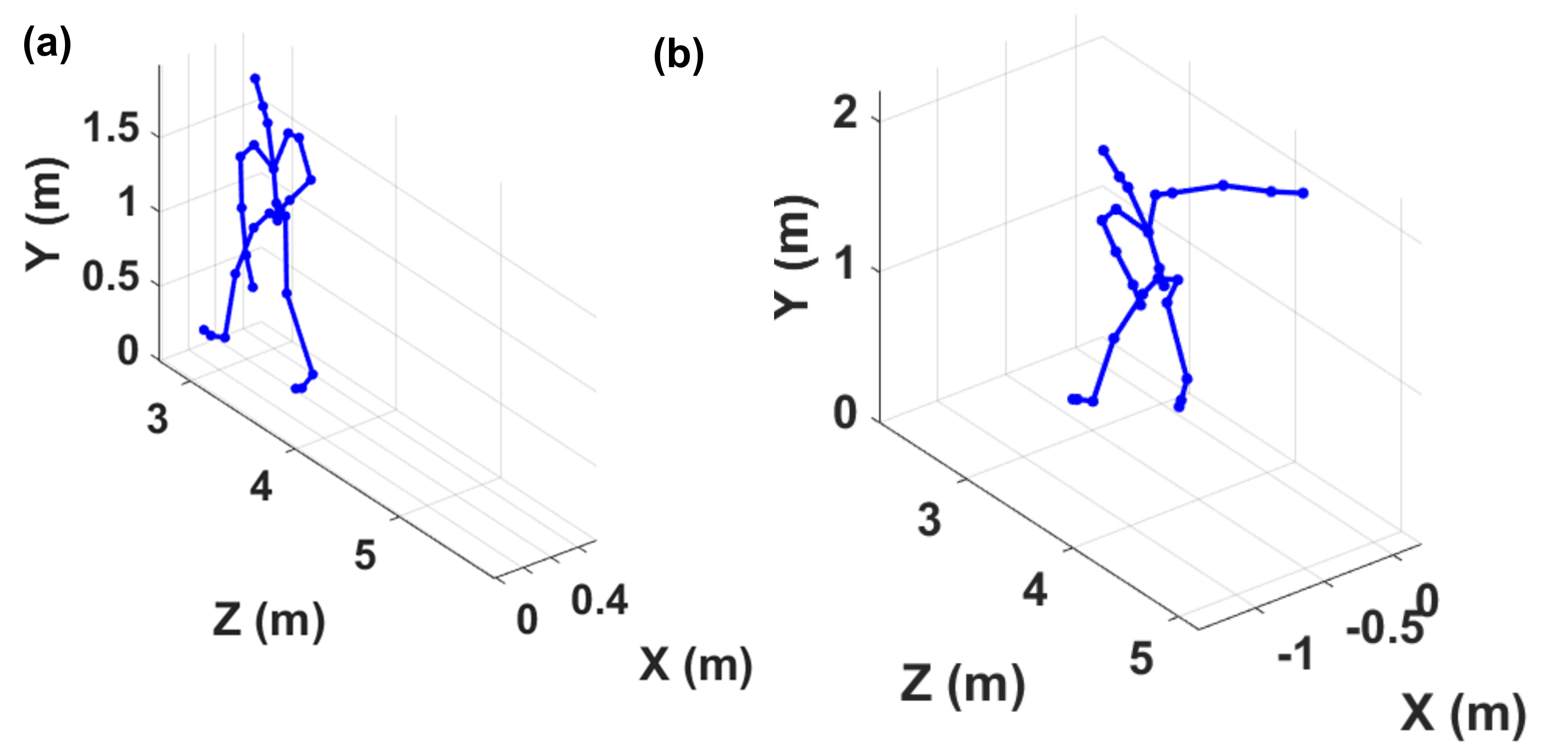}
\caption{Motion capture models of a human (a) walking (b) walk-leap-walk.}
\label{fig:human_motions}
\end{figure}

We consider the two human motion categories: (i) a natural human walking motion and (ii) a compound motion where the human first walks for 0.2 seconds, then leaps at 0.8 seconds and then resumes walking for the remaining 0.53 seconds. Each of these motions is of 1.53 s duration. For each of these motions, we consider four orientations of the human activities - 0, 15, 30, and 45 degrees w.r.t to the $x$-axis.

Each $b^{th}$ bone of the skeleton is modeled as an ellipsoidal body part. The human is modeled as a collection of $B$ discrete point scatterers corresponding to centroids of these body parts. The radar cross-section of each $b^{th}$ ellipsoid is given by $a_b^2$. The time-domain radar returns from the human, at frequency $f_c$, are obtained by
\begin{equation}
\label{eq:hybridem}
s_{rx}(t) = \sum_{b=1}^B A a_b \left(H(\vec{\rho}_b(t),f_c)\right)^2e^{-j4\pi \frac{f_c}{c} (r_b (t)- \rho_b(t)) }, 
\end{equation}
Here $r_b(t)$ is the time-varying Euclidean distance of the $b^{th}$ point scatterer from the radar, $\rho_b(t)$ is its projection in the 2D plane, and $c$ is the light speed. The two-way propagation physics from the radar to the point scatterer is captured by the square of the wall response $H$. $A$ calibrates the amplitude of the FDTD source excitation to the desired radar equivalent isotropic radiated power. The exponential term in the above expression maps the relationship between 2D through-wall propagation obtained from FDTD simulations and 3D free space propagation physics. 

Then we apply the STFT to the time domain radar data to obtain micro-Doppler spectrograms, $x_{DT}$, as shown in
\begin{equation}
    x_{DT}(t,f_D) = \int_{\tau} s_{rx}(\tau)w(\tau-t)e^{-j2\pi f_D \tau}d\tau, 
\end{equation}
where $w(t)$ is the short time window function of 0.2 s. The sampling frequency of the time domain data is 500Hz, resulting in Doppler frequency axes spanning from $f_D = -250Hz: +250Hz$ in all the spectrograms with a step size of 2.5Hz. Hence, the spectrogram size is $[239\times 200]$. 

\begin{figure}[htbp]
\centering
\includegraphics[width=90mm,height=65mm]{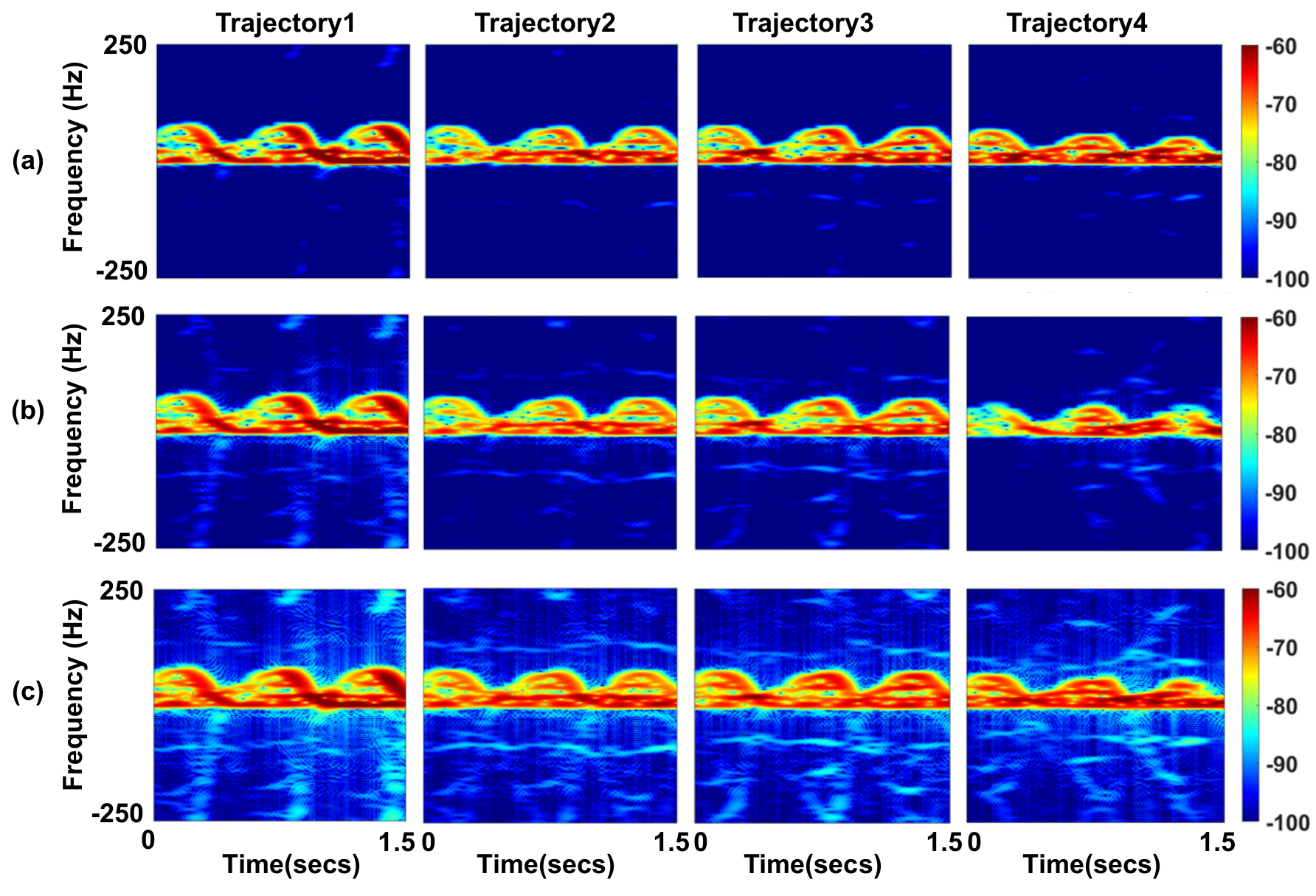}
\caption{Spectrograms of a human walk for four different orientations with respect to radar under (a) free space, (b) multi-layered wall, and (c) wall with air gaps scenarios.}
\label{fig:walk}
\end{figure}

In Fig.\ref{fig:walk}, the top row shows the spectrogram generated in free space conditions for the human walking motion for four orientations.
Here, we observe the strong returns from the torso and the weaker micro-Dopplers from the arms and legs. The second row depicts the spectrograms generated in multi-layered wall scenarios where the radar returns have undergone attenuation and multipath. The third row shows the effects of strong multipath generated within the wall with air gaps on the spectrograms. 
 
\begin{figure}[htbp]
\centering
\includegraphics[width=90mm,height=70mm]{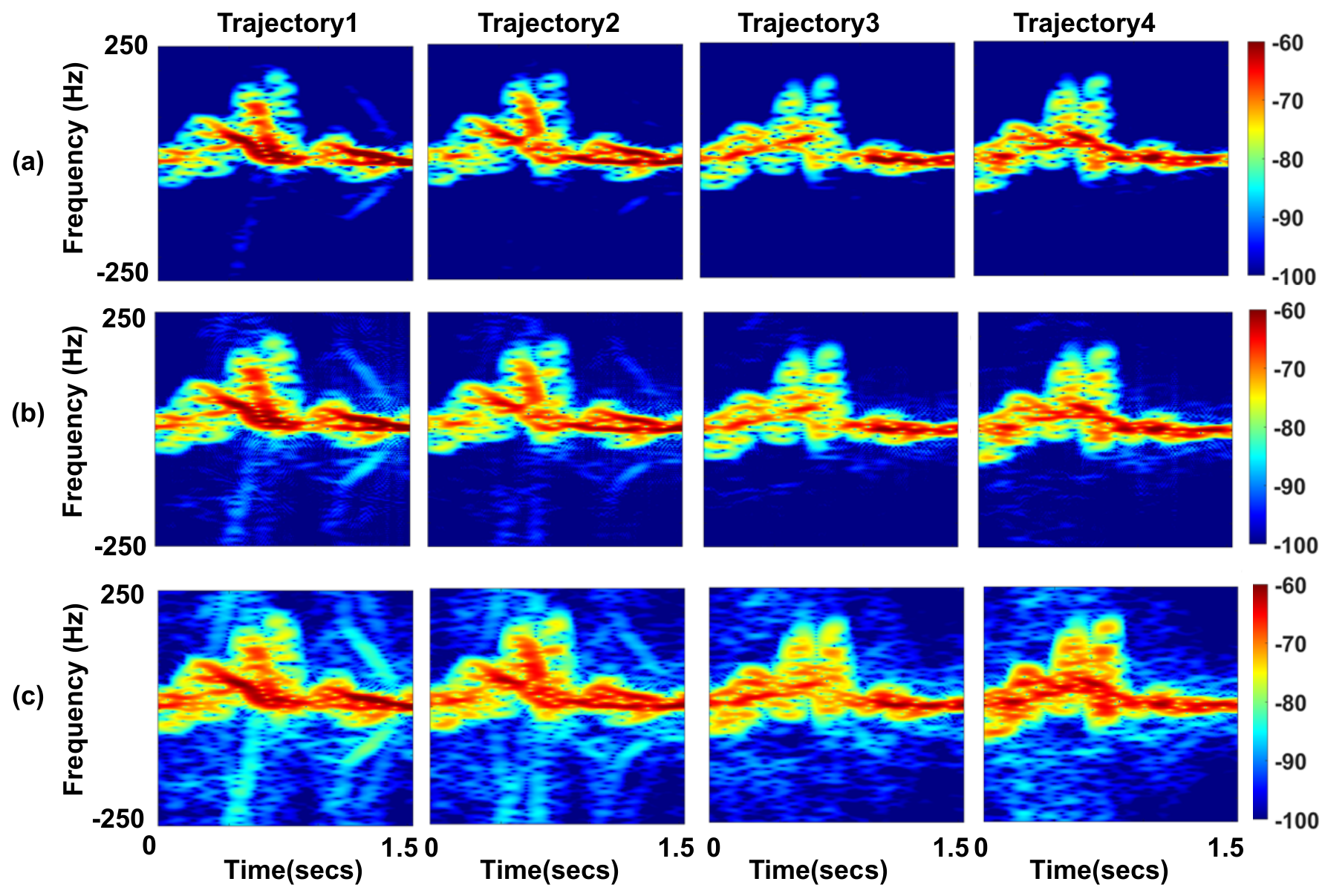}
\caption{spectrograms of walk-leap-walk human motion for four different orientations with respect to radar under (a) free space, (b) multi-layered wall, and (c) wall with air gaps scenarios.}
\label{fig:wlw}
\end{figure}

Similarly, we generated the spectrograms for walk-leap-walk for free space and through wall scenario as shown in Fig. \ref{fig:wlw}. 
Here, we see the steady micro-Dopplers of human walking motions followed by high-frequency components caused by the leap motion followed again by the walking patterns. From the through-wall radar micro-Doppler images in Fig.\ref{fig:wlw}b and Fig.\ref{fig:wlw}c, we observe that both the inhomogeneous walls introduce considerable attenuation and multipath to the spectrograms. 

We generated 960 through-wall micro-Doppler spectrogram samples as described in Table.\ref{table:simulated_data}. 
\begin{tiny}
   \begin{table}[!htbp]
\centering
\caption{Details of simulated radar database for training and testing the GAN}
\label{table:simulated_data}
\begin{tabular}{*5c}
\hline \hline 
\noalign{\vskip 1pt}
    {Type}&{sample}&{Aspect angles}&{Activities}&{total cases}\\
\hline 
\noalign{\vskip 1pt}
     {Free space}&{1}&{4}&{2}&{8}\\  
 {Multi-layered wall}&{60}&{4}&{2}&{480}\\
  {Wall with air gaps}&{60}&{4}&{2}&{480}\\
\hline \hline
\end{tabular}
\end{table} 
\end{tiny}
Out of the total samples, 80\% of the samples are used for training the GAN, and the remaining 20\% are retained for testing.  
\section{Results}
\subsection{Network Details}
In this section, we discuss the GAN network in detail. 
We consider a frequency span of the spectrogram from -160Hz to +160Hz due to the absence of noticeable components at higher frequencies. Additionally, we downsample the data by a factor of 2 to obtain 64-frequency samples. Further, we downsample the time-domain data to 64-time samples by a factor of 3. The resulting micro-Doppler spectrograms are of size $64\times64$. We consider the magnitude of these spectrograms as input to the GAN. The down-sampling operations in time and frequency are carried out to reduce the complexity of the deep learning architecture. Keras 2.7 is used to train and test the codes, which are executed on an Intel Core i7-10510U CPU clocked at 1.80 GHz.

The generator comprises an encoder and a decoder. The encoder consists of two convolutional transpose layers that upsample the input. The first layer has 32 filters, and the second layer has 64, both with a $2\times2$ kernel and stride. Both layers are followed by rectified linear unit (ReLU) activation functions and batch normalization, which add nonlinearity and stabilize training, respectively. The decoder section of the model comprises two convolutional layers that utilize ReLU activations and batch normalization, followed by max-pooling. The decoder also includes a convolutional transpose layer that uses a single filter with a kernel size of $4 \times 4$ and a stride of $1 \times 1$ with the sigmoid activation function. We provide the free space spectrogram of size $64\times64$ concatenated with Gaussian random noise of size $64\times64$ as input to the generator.

The critic network consists of three convolutional layers, each with ReLU activation functions. The first layer has 64 filters with a kernel size of $3\times3$ and a stride of $2\times2$. Subsequent layers include 128 and 256 filters with the same kernel and stride size. The output layer has
one node with a sigmoid function. We provide real through-wall input of size $64\times64$ to the critic.
\begin{figure}[htbp]
\centering
\includegraphics[scale=0.7]{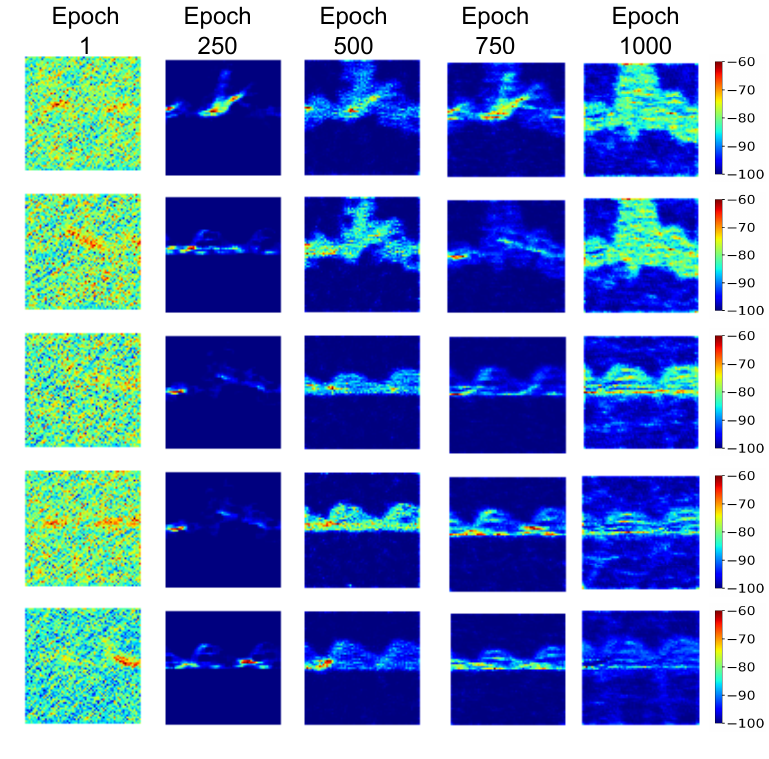}
\caption{Synthetic radar micro-Doppler spectrogram from five different free space micro-Doppler spectrogram input samples across the rows. The synthetic spectrograms generated from different epochs of the training algorithm are shown across columns. All the figures show Doppler frequencies spanning from -160Hz to +160Hz across the $y$ axis and 1.2 s across the $x$ axis.}
\label{fig:sample_training}
\end{figure}
The networks are configured with an Adam optimizer. The learning rate is set at 0.0002. The model is trained over 1000 epochs. The batch size is set to 64. A binary cross-entropy loss function measures the difference between the synthesized and real micro-Doppler spectrograms. 
 
\subsection{Results}
We evaluate the results both qualitatively and quantitatively. For visualization purposes, we show the synthesis of some samples of through-wall spectrogram across multiple epochs in Fig.\ref{fig:sample_training}.
This figure demonstrates how the neural network learns the wall effects on micro-Doppler spectrograms. The first and second rows show the training of walk-leap-walk motion, while the third, fourth, and fifth rows show the training of the walk motion. We observe that as the training progresses from epoch 250 to epoch 500, the network gradually gets better at synthesizing through-wall spectrograms and that by epoch 1000, the results are very similar to actual spectrograms obtained in through-wall conditions.

\begin{figure}[htbp]
\centering
\includegraphics[width=90mm,height=40mm]{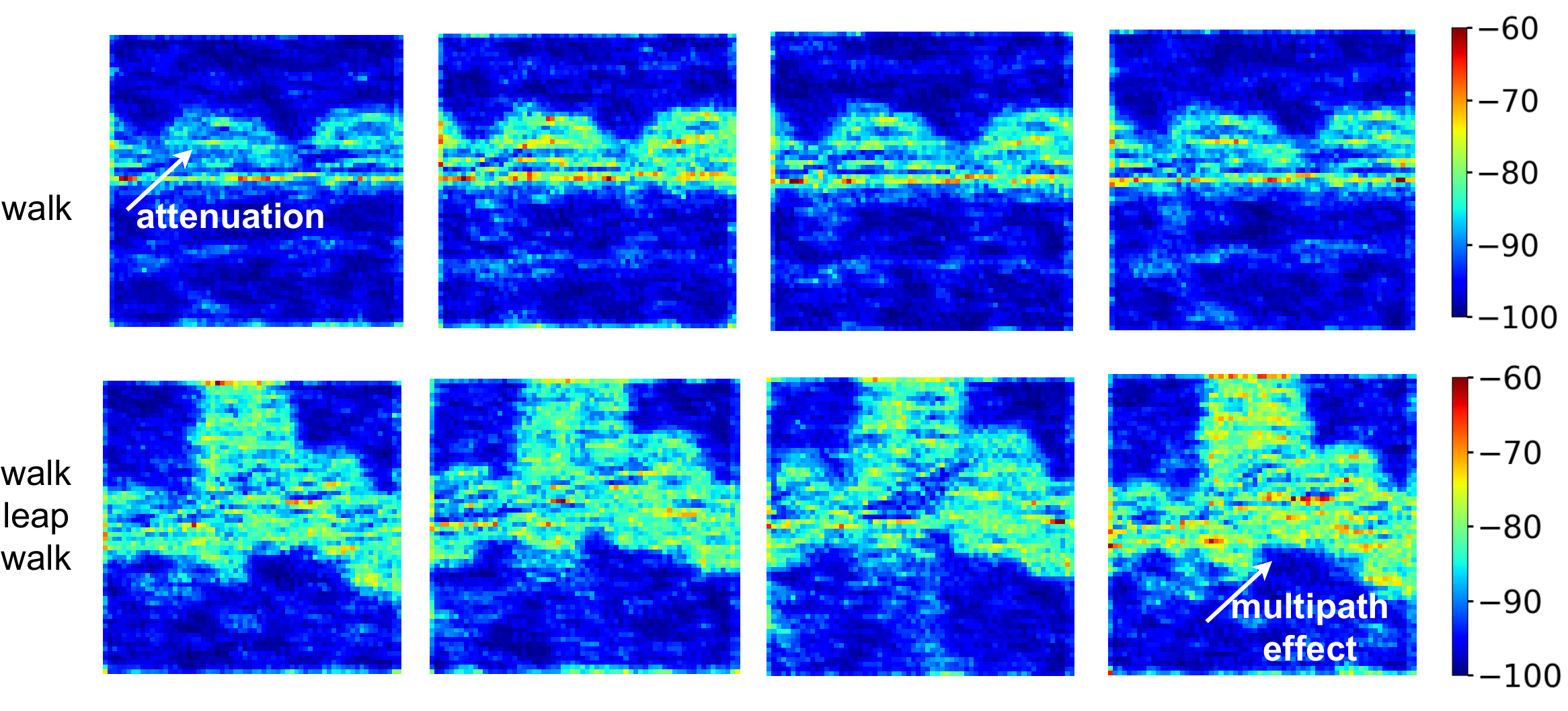}
\caption{Through-wall micro-Doppler spectrograms of human motions are generated from trained GAN. All the figures show Doppler frequencies from -160Hz to +160Hz across the $y$ axis and 1.2 s across the $x$ axis.} 
\label{fig:gen_img}
\end{figure}
Figure.\ref{fig:gen_img} shows examples of synthetic/fake through-wall samples generated from trained GAN for both types of human motions. In the first row, we
show four synthetic through-wall spectrograms of the human walk motion generated from an input of the free space counterparts at four orientations. The synthetic spectrograms show the micro-Doppler features of the torso, arms, legs, and multipath components. The second row of Fig.\ref{fig:gen_img} depicts through-wall spectrograms generated for walk-leap-walk for four different orientations. These figures also show that GAN successfully synthesized the micro-Doppler features and wall effects such as attenuation and multipath.
\begin{figure}[htbp]
\centering
\includegraphics[scale=0.45]{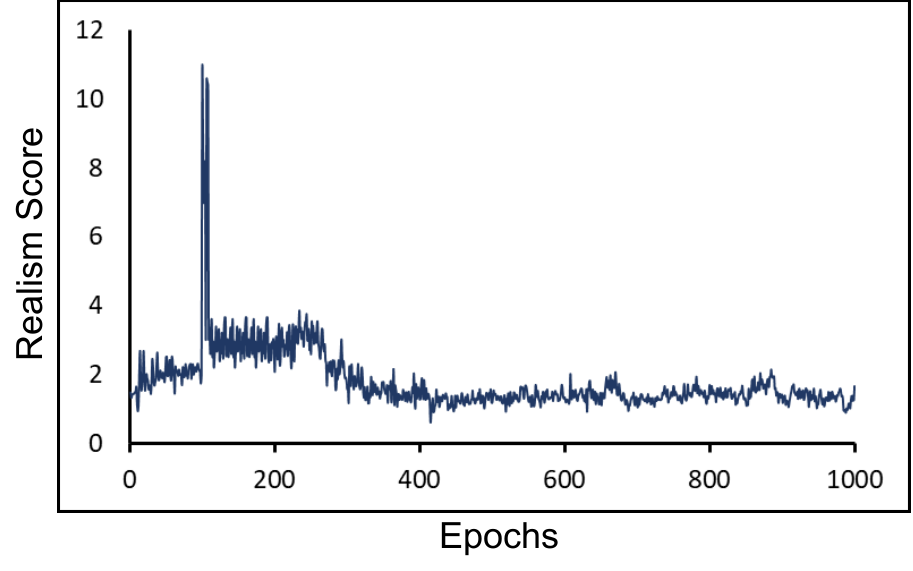}
\caption{Realism score for generated through-wall spectrograms obtained from the generator.}
\label{fig:realism_score}
\end{figure}

We study the realism of the generated through-wall images as a function of epochs in Fig. \ref{fig:realism_score}. The reconstruction error between the output and input of the DAE is falling with increasing epochs of the training process. By the $1000^{th}$ epoch, the error has fallen to around 1.6. 
As mentioned in Section II, this low reconstruction error indicates that the synthetic spectrograms generated from the GAN are realistic. 
\section{Conclusion}
Inhomogeneous walls introduce attenuation and multipath into radar micro-Doppler spectrograms of humans which affect the accuracy of classifiers. However, measurement and simulation data collection of human micro-Dopplers in through-wall environments is time-consuming and laborious.  
We have proposed using a GAN framework to synthesize through-wall micro-Doppler spectrograms from free space micro-Doppler spectrograms for different human motions, at different orientations and wall conditions. The synthetic micro-Doppler spectrograms show excellent realism. Future research directions include the generation of other types of commonly used through-wall radar signatures, such as range-Doppler and range-azimuth images; and the study of the effectiveness of classifiers trained on synthetic data in handling signatures/images generated from real radar measurements. 
\bibliographystyle{ieeetran}
\bibliography{main}

\end{document}